\documentclass[letterpaper,aps,pre,twocolumn,showcaps,superscriptaddress,amsmath,amssymb]{revtex4}
\usepackage{bmpsize}
\usepackage{graphicx,color}								
\usepackage{amssymb,amsmath}
\usepackage{natbib}
\usepackage{color}
\usepackage{array}
\usepackage{rotating}
\usepackage{subfigure}

\begin{document}

\author{Rhiannon Pinney}
\affiliation{HH Wills Physics Laboratory, Tyndall Avenue, Bristol, BS8 1TL, UK}
\affiliation{Bristol Centre for Complexity Science, University of Bristol, Bristol, BS8 1TS, UK}

\author{Tanniemola B.  Liverpool}
\affiliation{School of Mathematics, University of Bristol, Bristol, BS8 1TW, UK}

\author{C. Patrick Royall}
\affiliation{HH Wills Physics Laboratory, Tyndall Avenue, Bristol, BS8 1TL, UK}
\affiliation{School of Chemistry, University of Bristol, Cantock Close, Bristol, BS8 1TS, UK}
\affiliation{Centre for Nanoscience and Quantum Information, Tyndall Avenue, Bristol, BS8 1FD, UK}
\affiliation{Department of Chemical Engineering, Kyoto University, Kyoto 615-8510, Japan}

\email{paddy.royall@bristol.ac.uk}

\title{Structure in Sheared Supercooled Liquids: Dynamical Rearrangements of an Effective System of Icosahedra}

\begin{abstract}
We consider a binary Lennard-Jones glassformer whose super-Arrhenius dynamics are correlated with the formation of particles organized into icosahedra under simple steady state shear. We recast this glassformer as an effective system of icosahedra [Pinney \emph{et al. J. Chem. Phys.} \textbf{143} 244507 (2015)]. From the observed population of icosahedra in each steady state, we obtain an effective temperature which is linearly dependent on the shear rate in the range considered. Upon shear banding, the system separates into a region of high shear rate and a region of low shear rate. The effective temperatures obtained in each case show that the low shear regions correspond to a significantly lower temperature than the high shear regions. Taking a weighted average of the effective temperature of these regions (weight determined by region size) yields an estimate of the effective temperature which compares well with an effective temperature based on the global mesocluster population of the whole system. 
\end{abstract}


\maketitle

\section{Introduction}
\label{sectionIntroduction}

The mechanism behind the rapid dynamic slowing in liquids approaching the glass transition remains a mystery. There are many theoretical approaches to this problem, but a consensus on the nature of the liquid-to-glass transition is yet to be reached \cite{cavagna2009supercooled,berthier2011theoretical}. It has been proposed that icosahedral arrangements of the constituent atoms may form in some supercooled systems \cite{frank1952} and that dynamic arrest may be related to a (geometrically frustrated) transition to a phase of such icosahedra \cite{tarjus2005frustration,turci2016arxiv}. Geometric motifs such as icosahedra and other \emph{locally favoured structures} (LFS) can be identified in particle-resolved colloidal experiments \cite{konig2005,royall2008direct,hirata2009local,ivlev2012,leocmach2012,tamborini2015,zhang2016} and computer simulations \cite{dzugutov1992glass,coslovich2007understanding,eckmann2008ergodicity,sausset2010growing,tanaka2010critical,malins2013identification,royall2015strong,royall2016}. In particular, it has been shown that the onset of slow dynamics in simulated Lennard-Jones systems is closely coupled to the local structure, characterized by the LFS \cite{malins2013identification,speck2012first,hocky2014}.

A significant barrier to understanding the glass transition is its inaccessibility. Glassy systems have timescales that far exceed the practical limits of experimental or computational analysis \cite{berthier2011theoretical,royall2015physrep}. The operational glass transition is currently defined as the point when the liquid's viscosity exceeds a high enough value, \emph{i.e.} when the particles exhibit dynamic arrest on ``reasonable'' timescales \cite{berthier2011theoretical}. The temperature at which this happens is $T_{g}$. Direct detection of LFS and analysis of particle-resolved colloidal experiments and computer simulations are restricted to the first 4-5 decades of dynamic slowing, compared to 14 decades required to reach the operational glass transition ($T_{g}$) in molecular systems. Note that $T_g$ is distinct from lower temperatures at which the relaxation time of the material may diverge, such as that predicted by the Vogel-Fulcher-Tamman expression \cite{berthier2011theoretical,royall2015physrep}.

A complete picture of the glass transition therefore necessitates data extrapolation far below the accessible regime \cite{debenedetti2001supercooled}. Our previous publication \cite{pinney2015recasting} details how we have used the behavior of the LFS (icosahedra) to recast a well-studied binary Lennard-Jones glassformer into an effective system of LFS. To do so we have developed a population dynamics model of domains of icosahedra which we term \emph{mesoclusters}. Our model successfully describes the increase in relaxation time in terms of increasing mesocluster sizes and lifetimes as temperature is decreased and can be used to predict system behavior at significantly colder temperatures than those accessible to simulations. By construction, our model does not predict a thermodynamic phase transition to an ``ideal glass''.

In direct simulation and colloid experiments \cite{besseling2007}, a possible approach to probing deeper supercooling is to impose a shearing force on the system. Quiescent glasses exhibit dynamic heterogeneity; regions of high and low mobility. Shearing such an amorphous system can highlight the structural and dynamical subtleties that underlie glassy systems which may not have been otherwise observable. It has been shown that the local liquid-like (high mobility) regions can act as ``plasticity carriers'' \cite{demkowicz2005liquidlike} and shearing amorphous systems can allow the observation of some (otherwise elusive) long-range correlations in a colloidal glass \cite{chikkadi2011long}. In both experiments and computer simulations, locally ``soft'' and ``hard'' regions of the system, characterized by normal vibrational modes of inherent structures (\emph{soft modes}) \cite{widmer2009localized,candelier2009building,xu2010anharmonic,mosayebi2014soft}, configurational fluctuations that are susceptible to stress driven shear transformations (\emph{shear transformation zones, STZs}) \cite{falk1998dynamics,falk2010deformation,hassani2016} and localized regions of strong deformations (\emph{hot spots}) \cite{amon2012hot} have been shown to play a key role in the dynamics of supercooled liquids and the mechanics of amorphous solids. These can be used to predict when and where deformations will take place in sheared systems \cite{steif1982strain,rottler2014predicting,antonaglia2014bulk,mosayebi2014}. Recently, shear has been used to access the so-called Gardner transition \cite{charbonneau2014} between glass states with differing stabilities \cite{berthier2016,biroli2016}.

Imposing different shear rates can result in observing transitions between different states, such as: a continuous phase transition between brittle and hardening behavior \cite{dahmen2009micromechanical}, a dynamic transition between diffusive and arrested states \cite{fiocco2013oscillatory} and a first-order phase transition between banded and non-banded states \cite{varnik2003shear,greer2013shear,chikkadi2014shear}. Shear banding is the separation of a sheared system into two regions of different viscosity and internal structure \cite{martens2012spontaneous,chikkadi2014shear}. Some suggested mechanisms for the formation of shear bands are via the percolation of STZs \cite{ogata2006atomistic,antonaglia2014bulk} or from high stress localization in inherent defects or voids in the system \cite{shimizu2006yield}. Shearing has also been shown to increase the energy of soft glassy materials, called \emph{rejuvenation} \cite{bonn2002laponite,utz2000atomistic}, and varying the shear rate can yield systems with different \emph{effective temperatures} \cite{ono2002effective,berthier2002shearing,manning2009rate}. That is, increasing (decreasing) the shear rate is akin to increasing (decreasing) the temperature of the system.

Here we study the Wahnstr\"{o}m binary Lennard-Jones glassformer \cite{wahnstrom1991} under an imposed uniform planar shear. The LFS for this system was identfied as the icosahedron following an analysis of local environments of the constituent particles \cite{coslovich2007understanding}. Subsequently one of us investigated the lifetimes of 33 different structures, chosen to minimise the local potential energy \cite{doye1995,doye2003}. Of these, the icosahedron was found to last around a decade longer than other structures with a distinct bond topology \cite{malins2013identification}.

\begin{figure}
\begin{center}
\includegraphics[width=\columnwidth]{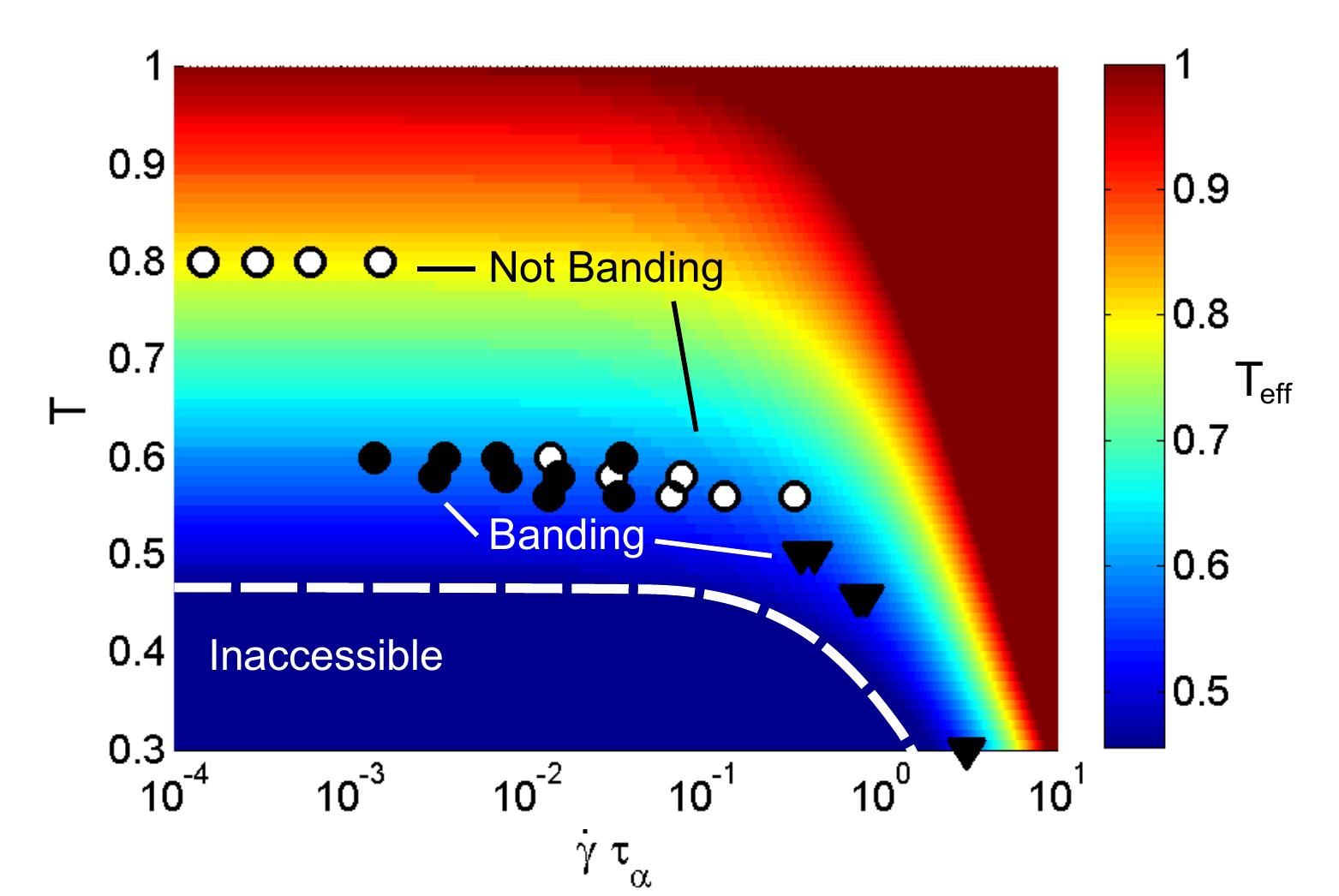}
\caption{State diagram for the Wahnstr\"{o}m model under shear. The effective temperatures obtained when systems with set simulation temperature $T$ are sheared with rates $\dot{\gamma}\tau_{\alpha}$ (Eq. \ref{eqShearTempConvergence}). Effective temperatures are shown as colour contours. Circular points indicate systems where $\tau_{\alpha}$ is directly calculable; triangular points are placed on the effective temperature contour corresponding to the simulation temperature $T$. Black points indicate banding, white points do not exhibit banding (see Eq. \ref{eqBandingCriteria} for criteria).}
\label{figEffectiveTemperaturePhaseDiagram}
\end{center}
\end{figure}

We find it is possible to obtain steady state behaviour at temperatures both above and below the glass transition temperature, which for our purposes is the temperature at which a Vogel-Fulcher-Tamman fit would diverge, $T_{\mathrm{VFT}} \approx 0.46$ (Fig. \ref{figEffectiveTemperaturePhaseDiagram}). At sufficient shear rates, temperatures which were inaccessible in quiescent simulations will reach a steady state and exhibit some characteristics, probed by structural properties of the LFS, typical of an effective temperature that is \emph{higher} than the actual simulation temperature. Previous attempts to define effective (or fictive) temperatures have used quantities such as free volume \cite{wright2003free}, energy \cite{lacks2004energy,langer2004dynamics} and interparticle forces \cite{corwin2005structural}. It is our aim to understand these sheared systems with an effective temperature determined by local structure, i.e. mesoclusters. Using the observed mesocluster properties in the sheared system and comparing them to our existing (temperature dependent) quiescent mesocluster model \cite{pinney2015recasting}, it is possible to determine the effective temperature of the sheared system. For systems that exhibit shear banding, we can determine the effective temperature of each region (high and low shear bands) using the same method. We find that increasing the shear rate results in an increase in effective temperature of the whole system and that the high and low shear bands have distinct effective temperatures; the high shear band has a significantly increased effective temperature.

This paper is organized as follows: we discuss the simulation protocol in Section \ref{sectionSimulation}. Section \ref{sectionGlobal} shows our effective temperature analysis for all simulations (all temperatures, all shear rates) looking at the ``global'' system; the system as a whole. Section \ref{sectionBanding} focuses on the systems that have exhibited shear banding where we study the high and low shear bands separately by cutting the simulation boxes into their relevant segments. We conclude with a summary and discussion in Section \ref{sectionSummary}.

\section{Simulation details} 
\label{sectionSimulation}

We simulate the Wahnstr\"{o}m equimolar binary Lennard-Jones model \cite{wahnstrom1991}. The size ratio is $5/6$ and the well depth between all species is identical. The mass of the large particles is twice that of the small. We use molecular dynamics simulations of $N=10976$ particles. We equilibrate for at least $100 \tau_\alpha$ in the NVT ensemble for $0.56 \leq T \leq 0.8$ and use the final configuration for $T = 0.56$ to initiate further NVT simulations at temperatures $0.3 \leq T \leq 0.5$ for as long as computationally possible. Here $\tau_\alpha$ is the structural relaxation time determined by a stretched exponential fit to the intermediate scattering function \cite{pinney2015recasting}.

The final configuration of each simulated temperature is used as the initial configuration of a sheared simulation following the SLLOD algorithm with Lees-Edwards periodic boundary conditions. All of these sheared simulations were carried out using LAMMPS \cite{LAMMPS}. The shear rates studied (in simulation units) are: $10^{-5} \leq \dot{\gamma} \leq 0.25$ for $0.56 \leq T \leq 0.8$ and $2.5 \times 10^{-6}$, $5 \times 10^{-6}$ and $10^{-5}$ for $0.3 \leq T \leq 0.5$. In our simulations, the yield point occurs at a strain $\gamma\approx0.1$. Here we take the steady state to correspond to $\gamma>1$ \cite{pinney2016transient}. We simulate up to strain values in excess of $\gamma = 2$, except in the case $T=0.3, \dot{\gamma} = 2.5 \times 10^{-6}$ where computational limits restrict the amount of strain simulated to $\gamma = 1.5$.

We identify icosahedra with the topological cluster classification (TCC) and consider those which last longer than $0.1\tau_\alpha$ (for $0.56 \leq T \leq 0.8$) or longer than 150 simulation time units (for $0.3 \leq T \leq 0.5$) to suppress the effects of thermal fluctuations. Here $\tau_\alpha$ is the structural relaxation time, determined from a fit to the intermediate scattering function \cite{pinney2015recasting}. Our structural analysis protocol is detailed in Ref. \cite{malins2013tcc}.

\section{Sheared systems: A global approach}
\label{sectionGlobal}

Shearing the system enough to reach a steady state (far beyond the yield point where steady stress is achieved) 
enables us to reach a steady state 
to temperatures that are otherwise inaccessible. By simulating a sheared system at an otherwise inaccessible temperature and modeling the mesocluster properties, it is possible to obtain a shear-rate dependent mesocluster model alongside the existing temperature-dependent mesocluster model. By combining two such models, we can more accurately predict the mesocluster properties (and thus the relaxation times) at temperatures approaching the glass transition. An overview of the results, and the state space accessible to the simulations, is shown in Fig. \ref{figEffectiveTemperaturePhaseDiagram}.

\subsection{Recap of population dynamics model}
\label{sectionRecap}

First, we briefly introduce the population dynamics model which generates the mesocluster size distribution from Ref. \cite{pinney2015recasting}. Mesoclusters are structures made up of particles in icosahedra, the LFS for the Wahnstr\"{o}m model glassformer \cite{coslovich2007understanding,malins2013identification}. We assume that mesoclusters of size $m$ ($m$ being the number of centres of icosahedra) can only change in size by $\pm 1$ and are restricted in size by a system-size dependent constant $M$. For high temperatures, $p_{m}$ (the probability of a mesocluster being size $m$) follows an exponential decay with steady-state solution 
\begin{equation}
p_{m}(T)= a(T)^{m-1}p_{1}(T) 
\end{equation}
where $a(T)$ is the temperature-dependent decay parameter. At lower temperatures, the mesoclusters percolate, and as such the shape of their size distribution changes. We account for this change by including a Gaussian weighting to obtain the steady state solution 
\begin{equation}
p_{m}(T) = a(T)W_m(T)p_{m-1}(T) 
\label{eqSteadyState}
\end{equation}
where $a(T)$ is an underlying decay parameter and $W_{m}(T)$ is the Gaussian weight which include ``mean'' and ``variance'' parameters to control the shape of the distribution. Our previous publication \cite{pinney2015recasting} discusses our mesocluster size model parameters in detail.

The mesocluster size distribution expected for a quiescent system at simulation temperature, $T$, may be described by Eq. \ref{eqSteadyState}. Supposing a sheared system exhibits mesocluster size distributions that are well described by this model (with no changes to the parameterization), we can conclude that at the level of our population dynamics model, the structure of the sheared system is similar to a quiescent system at model temperature $T$. Since the structure (characterized by LFS) and dynamics have been shown to be coupled in Lennard-Jones systems \cite{malins2013identification,pinney2015recasting,speck2012first}, we could expect the system dynamics of the sheared system to be similar to those of the quiescent system at model temperature $T$.

\subsection{The effect of shear on the mesoclusters}
\label{sectionEffectShear}

\begin{figure}
\begin{center}
\includegraphics[width=\columnwidth]{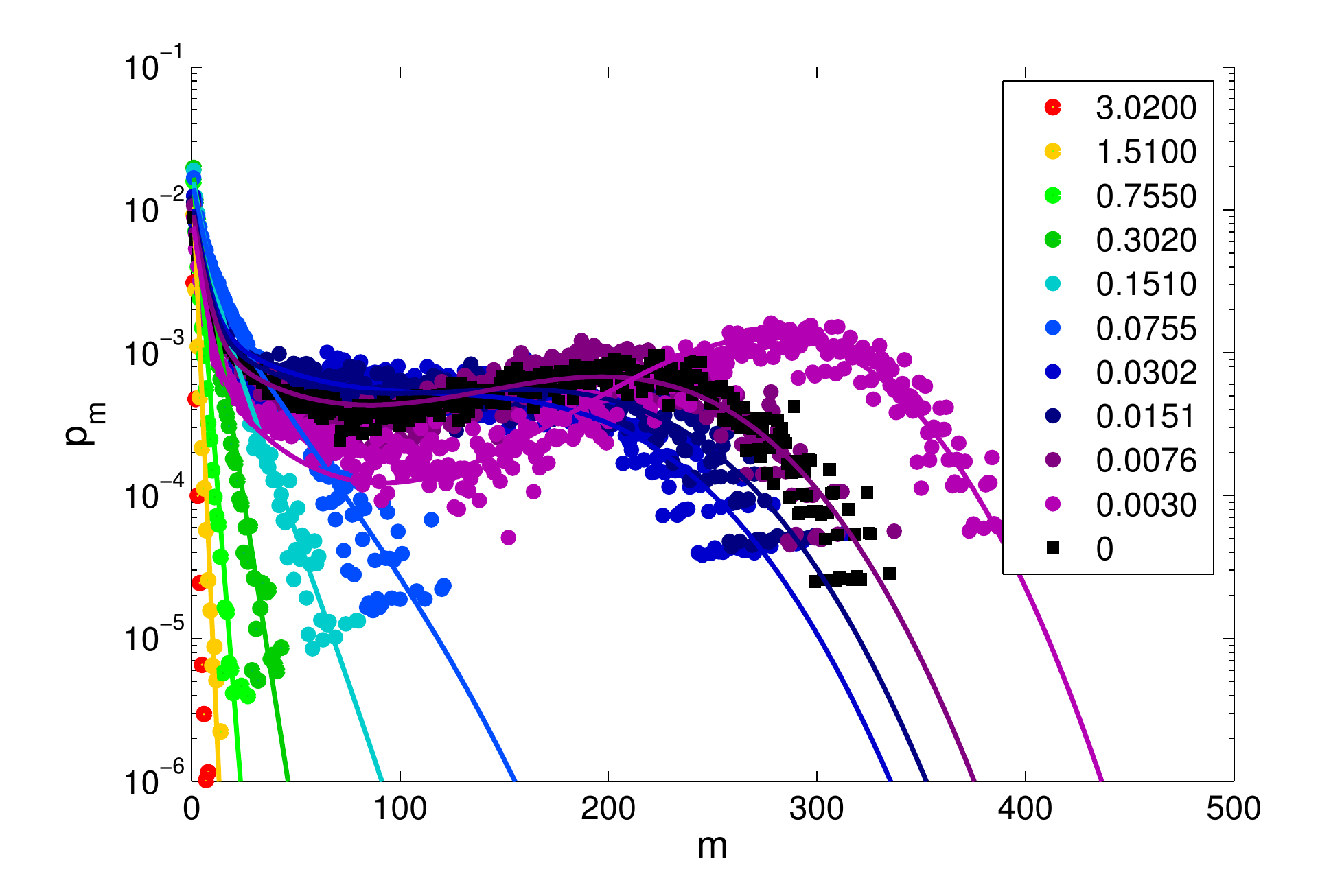}
\caption{The mesocluster size distribution for systems with $T = 0.58$ and varying shear rate, given in terms of $\tau_{\alpha}\dot{\gamma}$ in the legend for $0 \le \tau_{\alpha}\dot{\gamma} \le 3.02$.}
\label{figT058ShearProbs}
\end{center}
\end{figure}

We consider the changing mesocluster size distributions with varying shear rate for the full simulation box. Using the mesocluster size model as parameterized using the quiescent data from Ref. \cite{pinney2015recasting} (recalling Eq. \ref{eqSteadyState}), 
\begin{equation}
p_{m}(T) = a(T)W_m(T)p_{m-1}(T) \nonumber
\end{equation}
we can select a value of $T = T_{\mathrm{eff}}$ which results in the best fit of the model distribution to the observed mesocluster data for the sheared systems. It is this value of $T_{\mathrm{eff}}$ that we use as the effective temperature of the system. Figure \ref{figT058ShearProbs} shows the different mesocluster size distributions produced by varying the shear rate $\dot{\gamma}$ imposed on systems with $T = 0.58$. The mesocluster model distributions for the quiescent system (solid lines) are plotted alongside the simulation data for the system under shear. In each case, higher shear rates produce mesocluster size distributions typical of systems at higher temperatures.

\begin{figure}
\begin{center}
\includegraphics[width=\columnwidth]{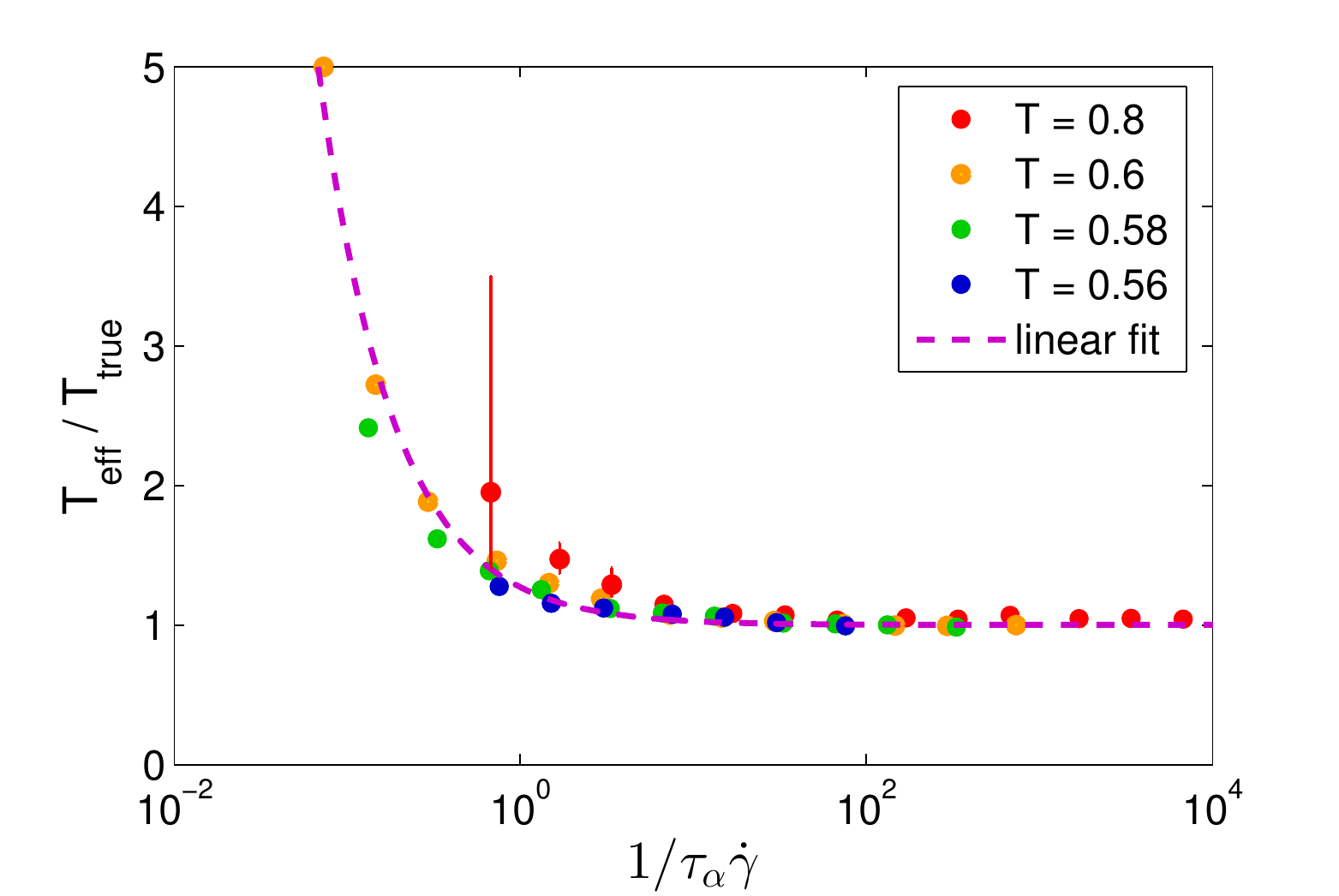}
\caption{As shear rates $\tau_{\alpha}\dot{\gamma}$ are decreased, the effective temperature $T_{\mathrm{eff}}$ converges to the ``true'' simulation temperature $T_{\mathrm{true}}$. For all temperatures, this happens at $\tau_{\alpha}\dot{\gamma} \lesssim 0.01$. Error bars are included on some $T = 0.8$ data points where the mesocluster statistics are limited due to low numbers of icosahedra and fitting the data is less constrained. Low temperature data, i.e. $T = 0.3$, is not included in this figure since $\tau_{\alpha}$ for such systems is not defined under our VFT fit with $T_0=0.46$.
}
\label{figShearTempLimits}
\end{center}
\end{figure}

For all systems, $T_{\mathrm{eff}} \rightarrow T_{\mathrm{true}}$ as the shear rate is decreased. Here $T_{\mathrm{true}}$ is the ``true'' simulation temperature.
Figure \ref{figShearTempLimits} shows the effective temperature in the sheared systems converging to the true simulation temperature. This can be fitted linearly using the following:
\begin{equation}
\frac{T_{\mathrm{eff}}}{T_{\mathrm{true}}} = 0.271 \dot{\gamma} \tau_{\alpha} + 1
\label{eqShearTempConvergence}
\end{equation}
At low shear rates $\tau_{\alpha} \dot{\gamma} < 0.01$ for any simulated temperature, $T_{\mathrm{eff}} \approx T_{\mathrm{true}}$.

\begin{figure}
\begin{center}
\includegraphics[width=\columnwidth]{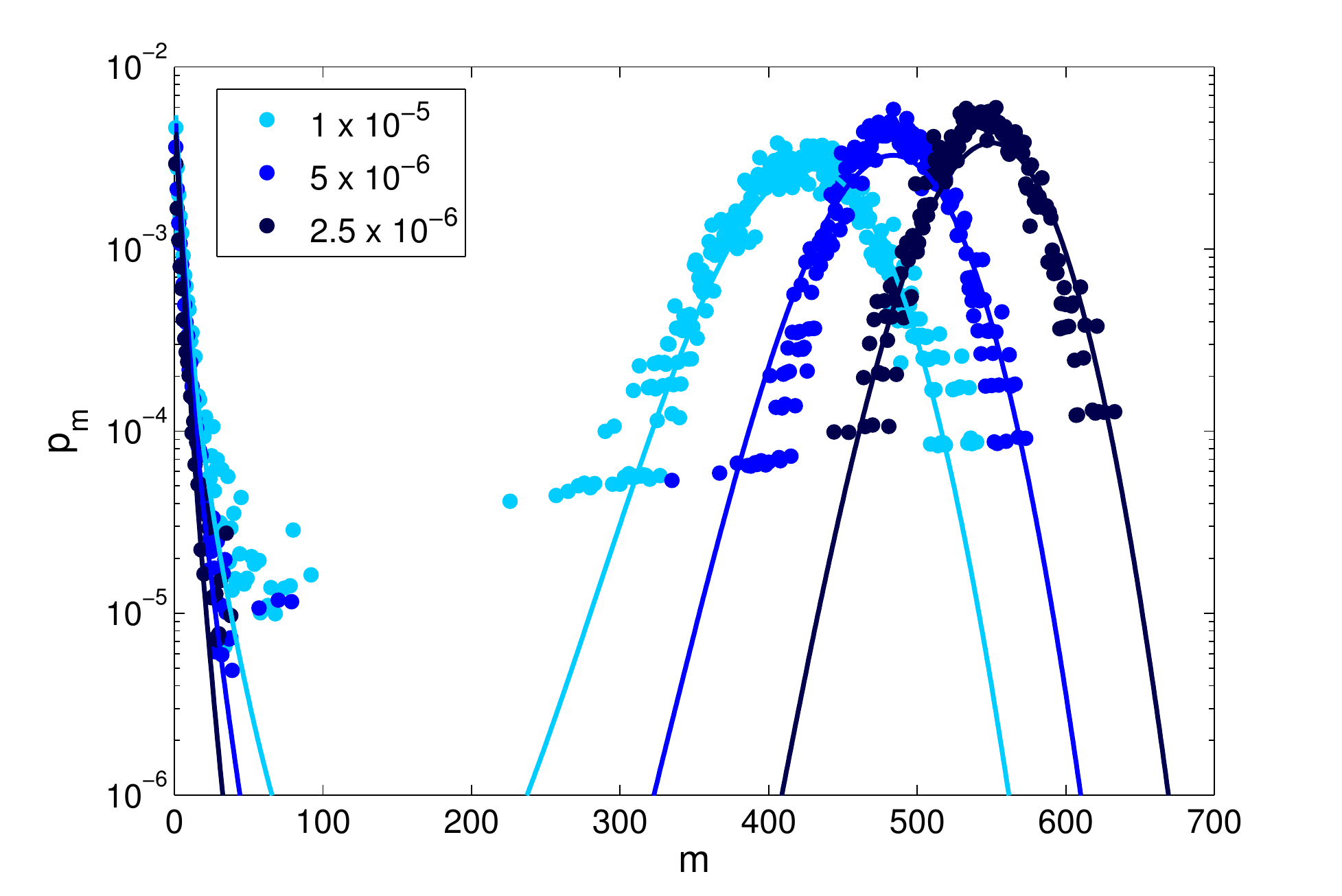}
\caption{The mesocluster size distribution for systems with $T = 0.3$ and varying shear rate, given in terms of simulation units in the legend. Fitted model lines correspond to $T = 0.556, 0.552, 0.548$, corresponding to a decrease in temperature as shear rate decreases.}
\label{figT03ShearProbs}
\end{center}
\end{figure}

Figure \ref{figT03ShearProbs} shows the mesocluster size distributions for $T = 0.3$; a significantly lower temperature than what is accessible in the quiescent regime, and in fact lower than $T_{\mathrm{VFT}}$. Thus the $\alpha$-relaxation time, $\tau_{\alpha}$, for this low temperature system is assumed to be infinite. Using the mesocluster size distributions from our population dynamics model \cite{pinney2015recasting}, we see that the effective temperatures of these sheared systems decrease as the shear rate is decreased, and are significantly colder than we have previously been able to access via quiescent systems ($T_{\mathrm{eff}} = 0.556, 0.552, 0.548$ in these sheared systems; quiescent systems are limited to $T \gtrsim 0.57$).

Across all temperatures and shear rates studied, the overall observed shape of the mesocluster distributions in the data sets and the model predictions are in excellent agreement with each other. Thus the data in the sheared systems can be accurately described by the mesocluster population model, and based on this observation we can assign an effective temperature to each. Furthermore, the deviation of the effective temperature from the true system temperature is linearly dependent on the rate of shear. In other words, \emph{within our mesocluster model shear rate and temperature can be superposed over one another.} This observation is made all the more remarkable by the fact that some of the state points we consider exhibit shear banding, which we now consider.  Figure \ref{figEffectiveTemperaturePhaseDiagram} shows the effective temperatures, $T_{\mathrm{eff}}$, of systems with varying \emph{simulation} temperatures, $T$, and shear rates following Eq. \ref{eqShearTempConvergence}.

\section{Shear Banding}
\label{sectionBanding}

\begin{figure}
\begin{center}
\includegraphics[width=\columnwidth]{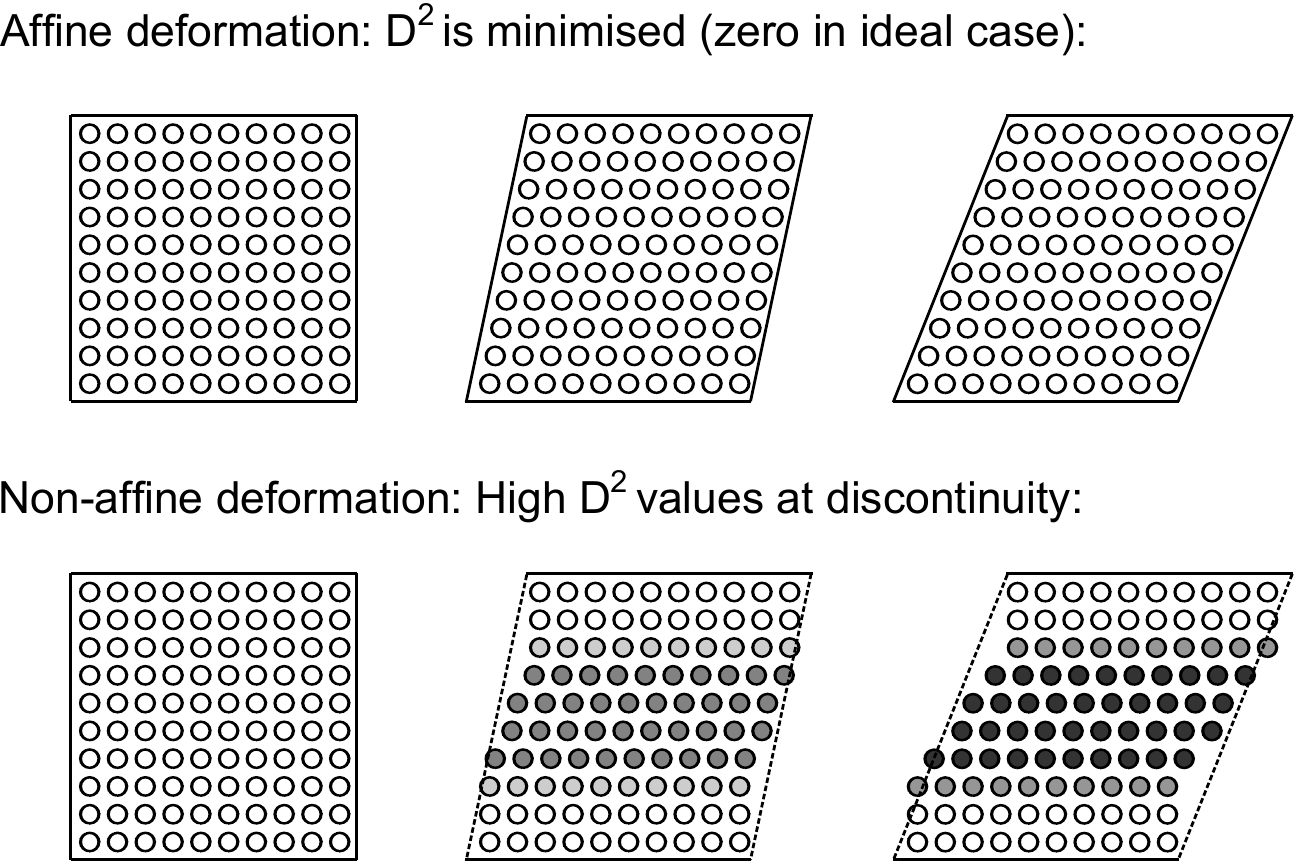}
\caption{A schematic of $D^2$ values in affine (top) and non-affine (bottom) displacements. For affine displacements, $D^2$ is uniformly minimized (equal to zero in the ideal case). In non-affine displacements, $D^2$ takes larger values located along the shear gradient discontinuity.}
\label{figD2schematic}
\end{center}
\end{figure}

So far, we have looked at the \emph{global} mesocluster properties of the sheared systems. However, these systems exhibit \emph{shear banding}, characterized in this case by a persistent $y$-axis dependence (perpendicular to the flow direction) in the icosahedra population and the corresponding local particle displacements measured using the non-affine deformation parameter, $D^{2}_{\mathrm{min}}$ \cite{falk1998dynamics}, and local shear rate. Figure \ref{figD2schematic} shows a schematic of $D^2_{\mathrm{min}}$ values expected in affine and non-affine displacements. Equation \ref{eqD2} is the definition of $D^2_{\mathrm{min}}$ as given in Ref. \cite{gannepalli2001molecular}, where $N$ is the number of neighbouring particles within the interaction range of a central particle, and the positions of the central particle, $n = 0$, and neighbouring particles, $n \in [1,N]$, given by $\textbf{r}_{n}(t)$ and $\textbf{r}_{n}(\tau)$ at times $t$ and $\tau = t - \Delta t$ respectively. We henceforth drop the subscript ``min'' for ease of notation in later equations.
\begin{equation}
D^{2}(\tau, t) = \sum_{n=1}^{N} \textbf{R}_{n} \cdot \textbf{R}_{n}^{T}
\label{eqD2}
\end{equation}

\begin{align}
\textbf{R}_{n} &= \Big(\textbf{r}_{n}(t) - \textbf{r}_{0}(t) \Big) - \Big(\textbf{XY}^{-1} \Big) \cdot \Big(\textbf{r}_{n}(\tau) - \textbf{r}_{0}(\tau) \Big) \nonumber \\
\textbf{X} &= \sum_{n=1}^{N} \Big(\textbf{r}_{n}(t) - \textbf{r}_{0}(t) \Big) \Big(\textbf{r}_{n}(\tau) - \textbf{r}_{0}(\tau) \Big) \nonumber \\
\textbf{Y} &= \sum_{n=1}^{N} \Big(\textbf{r}_{n}(\tau) - \textbf{r}_{0}(\tau) \Big) \Big(\textbf{r}_{n}(\tau) - \textbf{r}_{0}(\tau) \Big)
\label{eqD2extras}
\end{align}

\begin{figure*}
\begin{center}
\includegraphics[width=180mm]{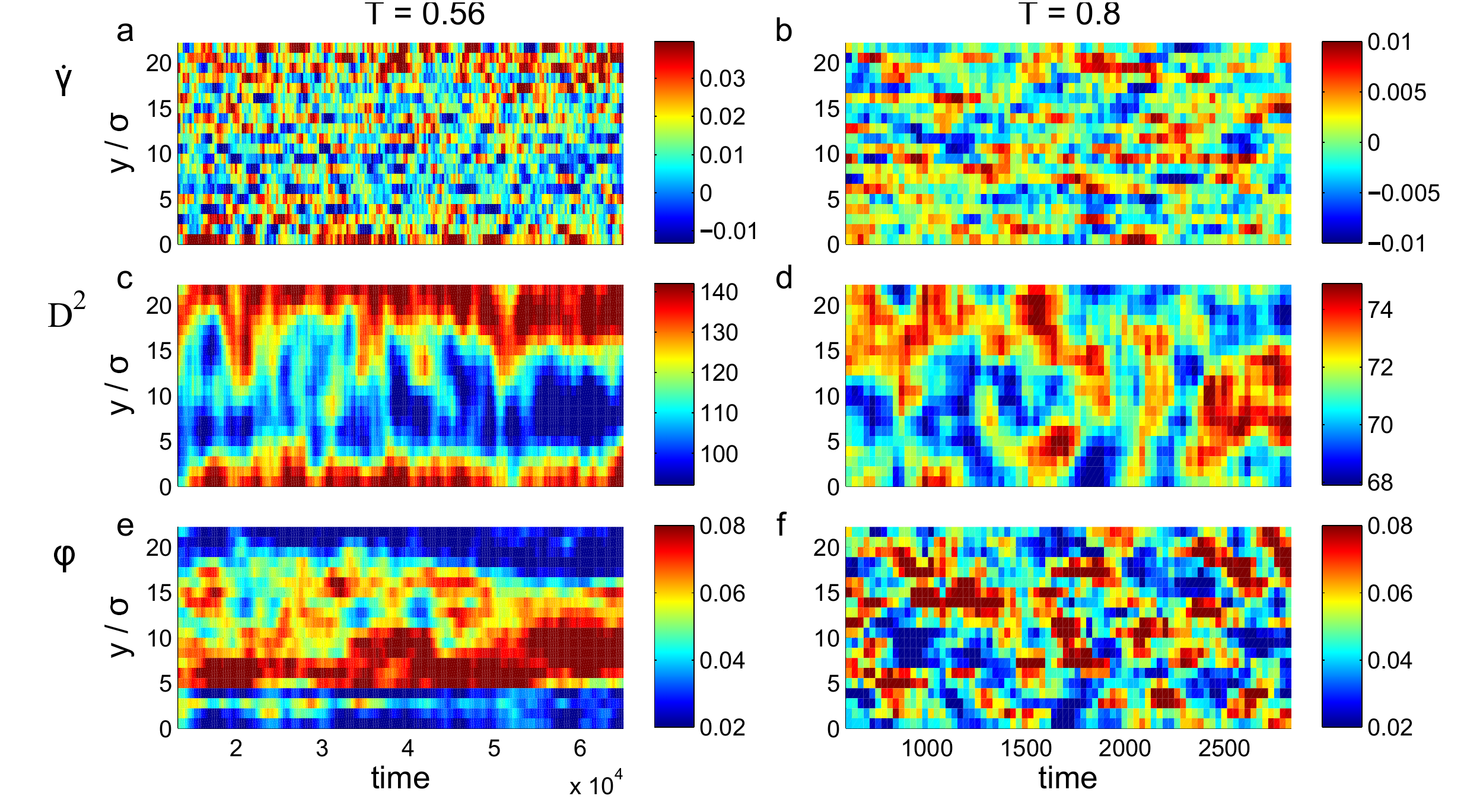}
\caption{Time-evolution of shear rate for each $y$-axis region given in terms of particle diameters $\sigma$ from the bottom of the $y$-axis, (a,b), non-affine dynamics (c,d) and population of icosahedra (e,f) for selected state points. Left column: $T = 0.56, \tau_{\alpha} \dot{\gamma} = 0.0132$ exhibits banding. Right column: $T = 0.8, \tau_{\alpha} \dot{\gamma} = 5.91 \times 10^{-5}$ does not exhibit banding.}
\label{figBandingExample}
\end{center}
\end{figure*}

Figure \ref{figBandingExample} shows examples of banded and non-banded sheared systems, distinguished by the values of shear rate, $D^{2}$ and the relative density of icosahedra in each binned region of the $y$-axis for progressing simulation time. We see that $D^2$ provides a clearer interpretation than the local shear rate.

\begin{figure}
\begin{center}
\includegraphics[width=\columnwidth]{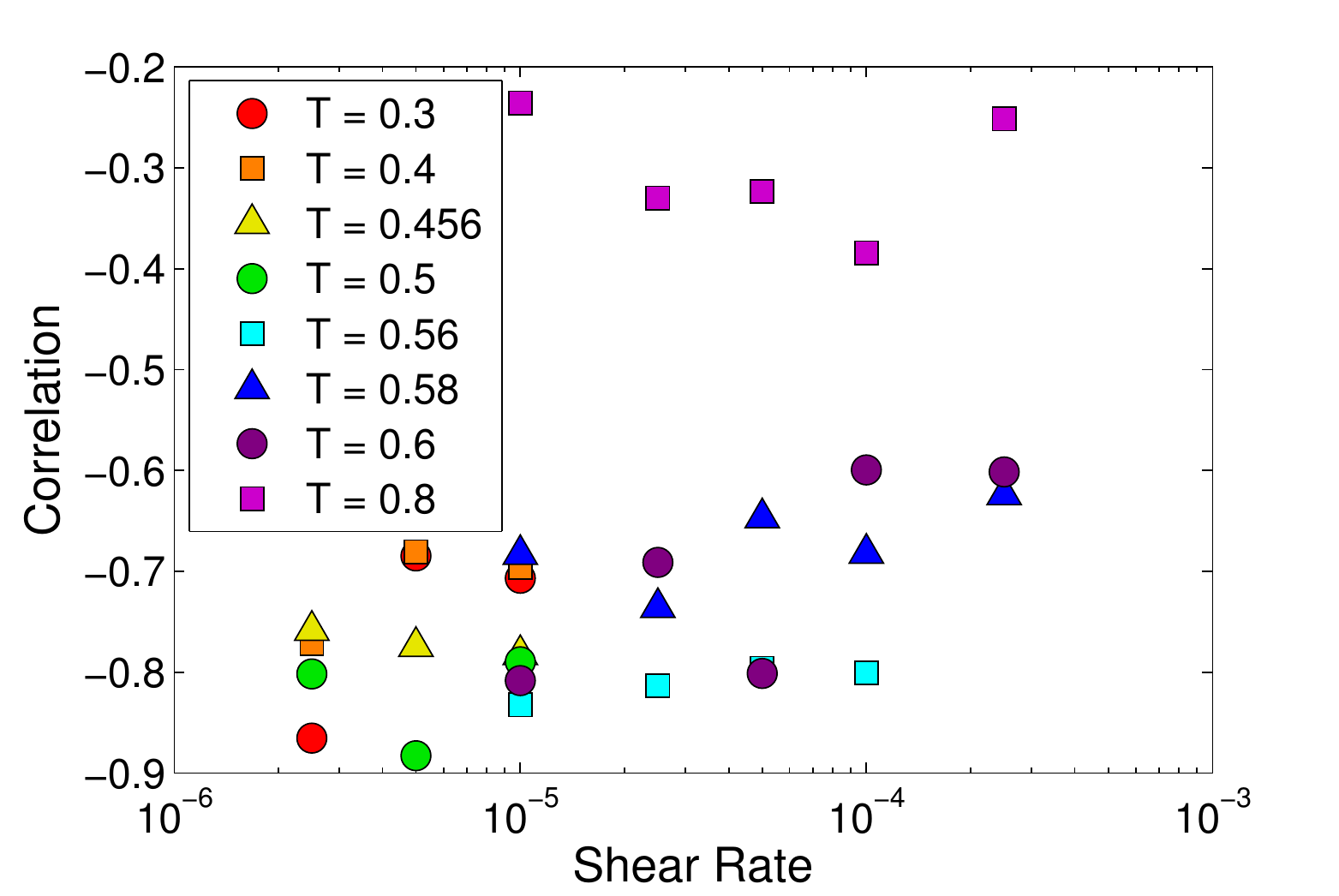}
\caption{Correlation coefficients for state points with varying $T$ and $\dot{\gamma}$ (simulation units). For comparison: correlation coefficients for quiescent state points are -0.58, -0.6, -0.5 and -0.38 for $T =$ 0.6, 0.65, 0.7 and 0.8 respectively.}
\label{figCorrelations}
\end{center}
\end{figure}

In Fig. \ref{figBandingExample}, we consider two representative temperatures, $T=0.56$ and $T=0.8$. In the former case, the product of the shear rate and the structural relaxation time $\tau_{\alpha} \dot{\gamma} = 0.0132$: here both the structure (in terms of the population of icosahedra) and the shear band are long lived. Conversely, at the higher temperature, $ \tau_{\alpha}\dot{\gamma} = 5.91 \times 10^{-5}$ and no banding is observed. Figure \ref{figCorrelations} shows the total correlation coefficient between $D^{2}$ and the density of icosahedra for all $T$ and $\dot{\gamma}$ obtained from data such is those shown in Fig. \ref{figBandingExample}. In banded systems, the correlation coefficient is strongly negative. The correlation between shear rate and $D^2$ across the binned regions of the $y$-axis is strong in systems where banding is exhibited. This suggests that there may be a causal relationship between the icosahedra dense regions of the system and the slow shear bands (this will be investigated in a future publication \cite{pinney2016transient}).

\begin{figure}
\begin{center}
\includegraphics[width=\columnwidth]{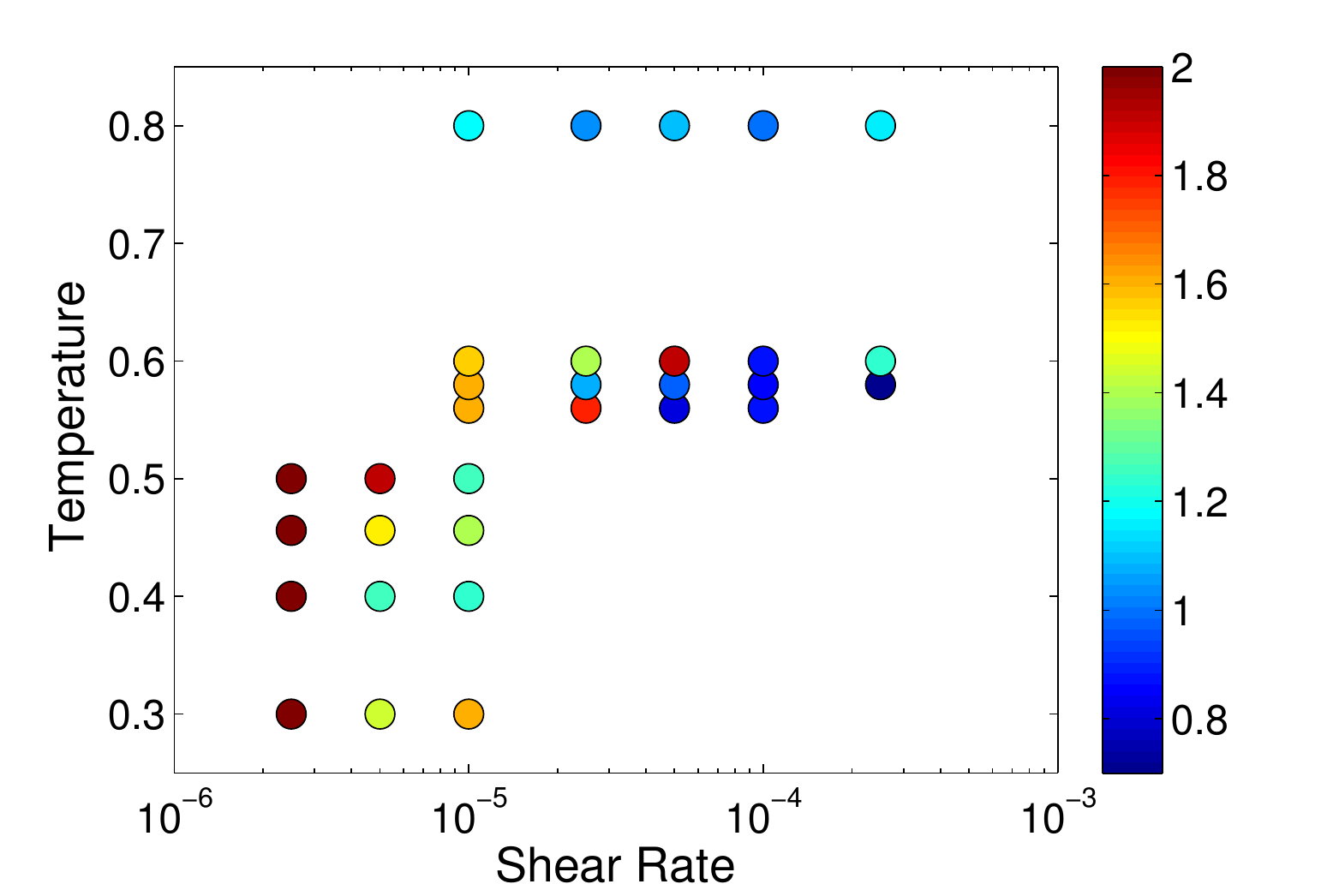}
\caption{The values of the banding criterion $R$ (Eq. \ref{eqBandingCriteria}) over a range of temperatures and shear rates. Deep blue indicates no banding and orange/red indicate very strong, persistent banding. Intermediate colours indicate systems which may exhibit a mix of behaviours through time, suggesting weak or intermittent banding.}
\label{figBandingPhaseDiagram}
\end{center}
\end{figure}

The simulation box was segmented along the $y$-axis to form 20 equal bins of roughly 1 particle diameter in height. Each bin is characterized by the average $D^2$ value of all the particles residing within that bin. To quantify whether or not a system is banding, we compare the average range of different $D^2$ values observed across the $y$-axis with the average range of $D^2$ values observed within each $y$-axis slice through time:
\begin{equation}
R = \frac{\left\langle D^{2}_{\mathrm{max}} - D^{2}_{\mathrm{min}} \right\rangle_{y}}{ \left\langle D^{2}_{\mathrm{max}} - D^{2}_{\mathrm{min}} \right\rangle_{t}}
\label{eqBandingCriteria}
\end{equation}
\noindent
where subscripts $y,t$ are the parameters to be averaged over; $y$-axis and time respectively. The value of $R$ quantifies how strongly banded the system is. Strong banding is characterized by large values of $R$. Systems which appear to fluctuate between banding and not banding through time have values $0.9 \lesssim R \lesssim 1.1$. $R < 0.9$ suggests that the system is not banding at all. Figure \ref{figBandingPhaseDiagram} shows the resulting values of $R$ for a number of state points.

\subsection{Identifying the shear bands}
\label{sectionIdentifying}

\begin{figure}
\begin{center}
\includegraphics[width=\columnwidth]{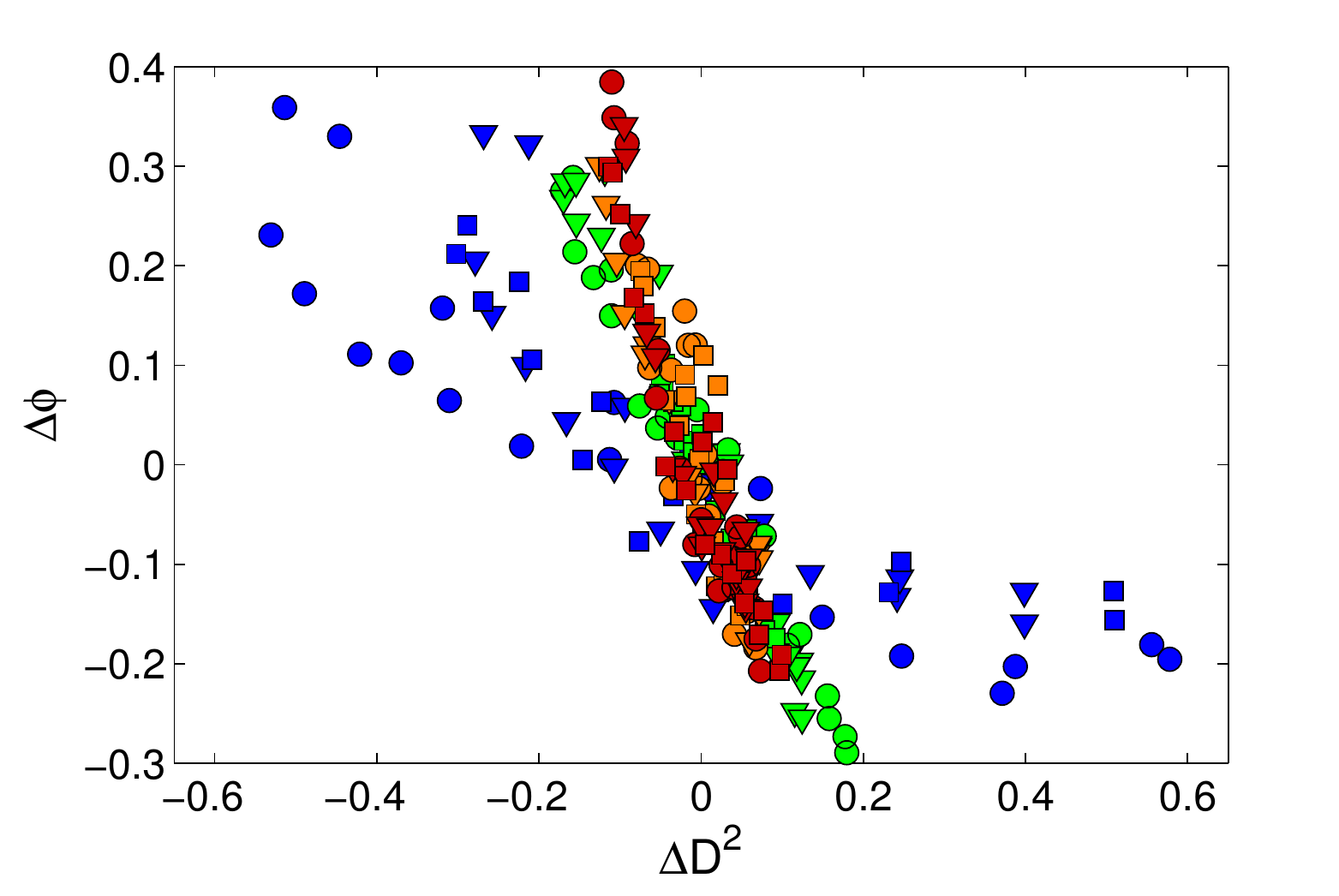}
\caption{Blue: $T = 0.3$. Green: $T = 0.56$. Yellow: $T = 0.58$. Red: $T = 0.6$. Circles: $\dot{\gamma} = 1 \times 10^{-5}$, triangles: $\dot{\gamma} = 2.5 \times 10^{-5}$, squares: $\dot{\gamma} = 5 \times 10^{-5}$. Blue triangles: $\dot{\gamma} = 5 \times 10^{-6}$, blue squares: $\dot{\gamma} = 2.5 \times 10^{-6}$.}
\label{figD2phiMapping}
\end{center}
\end{figure}

Since the population of icosahedra has a $y$-axis dependence in the shear banded systems, it should be the case that the mesocluster distributions vary across these regions. The average non-affine deformation, $D^{2}$, and the proportion of particles in icosahedra, $\phi$, values were calculated for each $y$-axis bin. From these we obtain local deviations in $D^{2}$ and $\phi$ as follows:
\begin{align}
\Delta D^{2} &= \frac{D^{2} - \bar{D}^{2}}{\bar{D}^{2}} \nonumber \\
\Delta \phi &= \frac{\phi - \bar{\phi}}{\bar{\phi}}
\end{align}
\noindent
$\Delta D^{2}$ and $\Delta \phi$ are plotted against each other in Fig. \ref{figD2phiMapping}. From this plot, we can see that the proportion change in $\phi$ can be determined (and predicted) from a linear mapping of the proportion change in $D^{2}$. Note that this linear mapping passes through the origin; \emph{i.e.} no change in $D^{2}$ means no change in $\phi$.

The banded systems exhibit one low shear region (low $D^2$, high icosahedra density) and one high shear region (high $D^2$, low icosahedra density). We can quantitatively define these regions by using the $D^2$ values. The simplest way of separating these regions would be to cut along the average $D^2$ value and look at the above (below) average segments. However, the $D^2$ values do not show any sharp transition from above (below) average. Instead, they smoothly increase (decrease) over the $y$-axis bins, thus blurring the exact boundary locations between the two regions. For this reason, we partition the simulation box into three types of regions: high shear, low shear and ``interface''. These interface regions are often small (the system is dominated by the high and low shear regions) and likely to be non-trivial combinations of low and high shear behavior, however we would expect them to behave approximately as an ``average'' between high and low. Since we are mainly interested in the behavioral differences between the high and low shear regions, we will focus only on these segments. The different segments can be defined using the following boundary definitions:
\begin{align}
S_{h} &= D^{2}_{\mathrm{av}} + \frac{D^{2}_{\mathrm{max}} - D^{2}_{\mathrm{av}}}{A} \nonumber \\
S_{l} &= D^{2}_{\mathrm{av}} - \frac{D^{2}_{\mathrm{av}} - D^{2}_{\mathrm{min}}}{A} 
\label{eqBoundaries}
\end{align}
\noindent
where $A$ is a number which can be chosen to increase or decrease the size of the interface regions (we set $A = 2$), and $S_{h} (S_{l})$ represents the lower (upper) $D^{2}$ boundary value of the high (low) shear segment.

\subsection{Mesoclsuster sizes in the bands}
\label{sectionMescoclusterSizesBands}

Once the different shear rate segment locations have been determined, mesocluster size analysis can be carried out on each segment individually. Now the mesocluster size model is system size dependent, due to the effect of the percolating mesocluster upon the size distribution. We have previously determined suitable parameters for the model for $N=1372,10976$ and $87808$ \cite{pinney2015recasting}. These we interpolate here, noting that the model parameters were obtained for cubic systems. This is achieved by using the same methods developed in our previous publication \cite{pinney2015recasting}, but only considering icosahedra whose \emph{centres} reside inside the segment, and only counting the particles in icosahedra which lie inside the segment. This will result in some partial icosahedra along the boundary edges, but is the simplest method of partitioning the simulation box and its mesoclusters.

Further to this, since the exact location and height of the shear bands vary slightly through time, the simulations are split into 8 equal time windows (i.e. $400\tau_{\alpha}$ is split into $8 \times 50\tau_{\alpha}$ time windows) and the segment boundaries defined for each. When defined with the interface parameter $A = 2$ in Eq. \ref{eqBoundaries}, most of the low/high shear segments are $\geq20\%$ of the height of the simulation box, which can all be reasonably described with system size dependent mesocluster size models. These system size dependent models can become somewhat inaccurate in describing the observed mesocluster size distributions in thinner segments. A handful of the high and low shear segments fall below this threshold, but they are infrequent enough not to cause significant effects in the results. Generally, the high shear segments are $\approx 20-25\%$ of the simulation box height and the low shear segments are $\approx 40-50\%$ of the simulation box height.

\begin{figure}
\begin{center}
\includegraphics[width=\columnwidth]{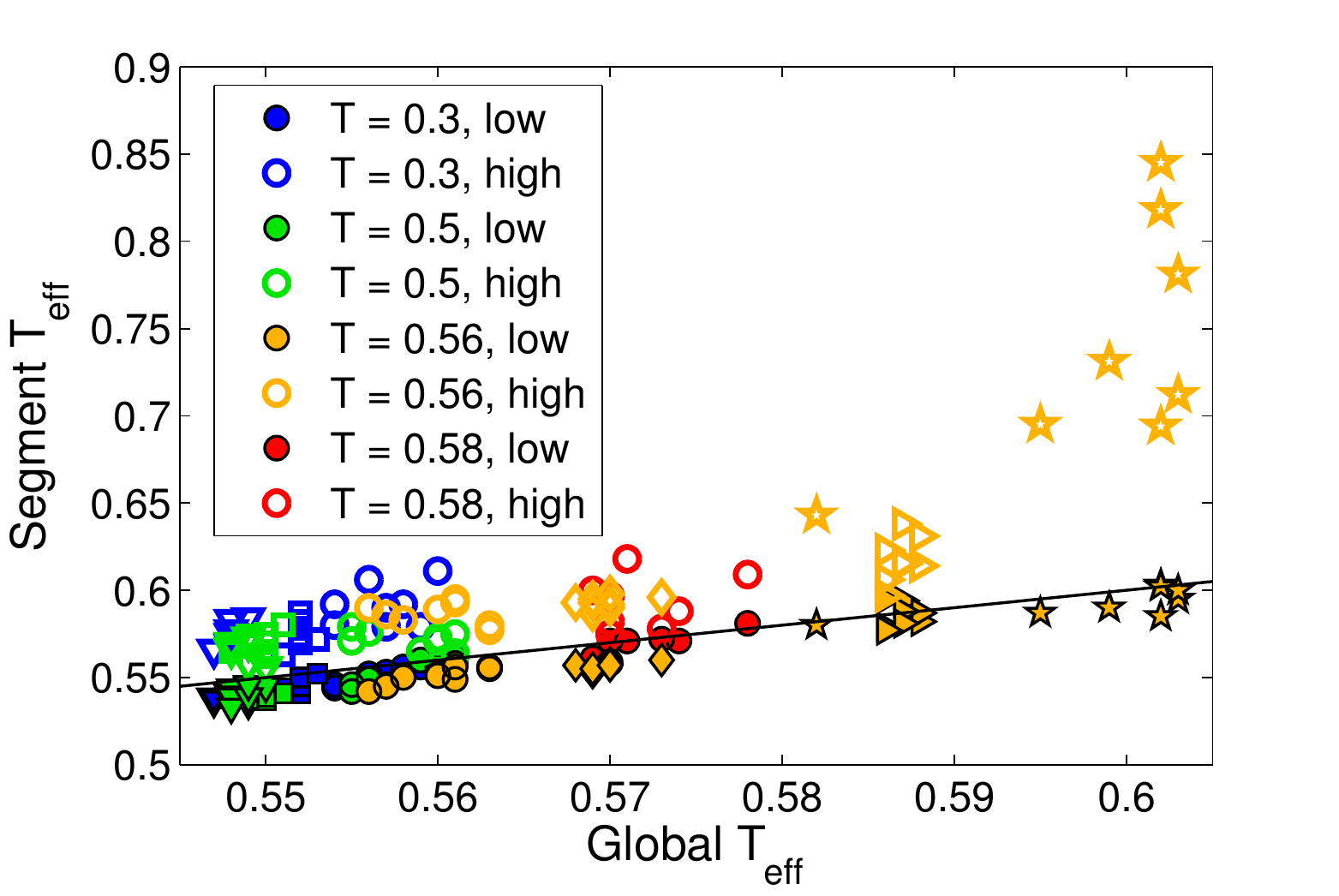}
\caption{The effective temperatures of the segments plotted against the global effective temperatures. The low (filled) and high (unfilled) shear segments display effective temperatures that are (respectively) below and above the global values.
Different shear rates are denoted by the shapes of the symbols: 
triangle down: $\dot{\gamma} = 2.5 \times 10^{-6}$; 
square:  $\dot{\gamma}= 5 \times 10^{-6}$; 
circle:  $\dot{\gamma}= 10^{-5}$; 
diamond:  $\dot{\gamma}= 2.5 \times 10^{-5}$; 
triangle right:  $\dot{\gamma}= 5 \times 10^{-5}$; 
star: $\dot{\gamma}= 10^{-4}$; 
}
\label{figSegmentTemperatures}
\end{center}
\end{figure}

Figure \ref{figSegmentTemperatures} shows the fitted effective temperatures of the high and low shear segments for different temperatures and shear rates. The low shear segments have effective temperatures lower than the global averages (calculated as in Section \ref{sectionGlobal} for the 8 time windows), and the high shear segments have significantly higher effective temperatures. This is mirrored in the observed values of $\phi$ across these segments.

\begin{figure}
\begin{center}
\includegraphics[width=\columnwidth]{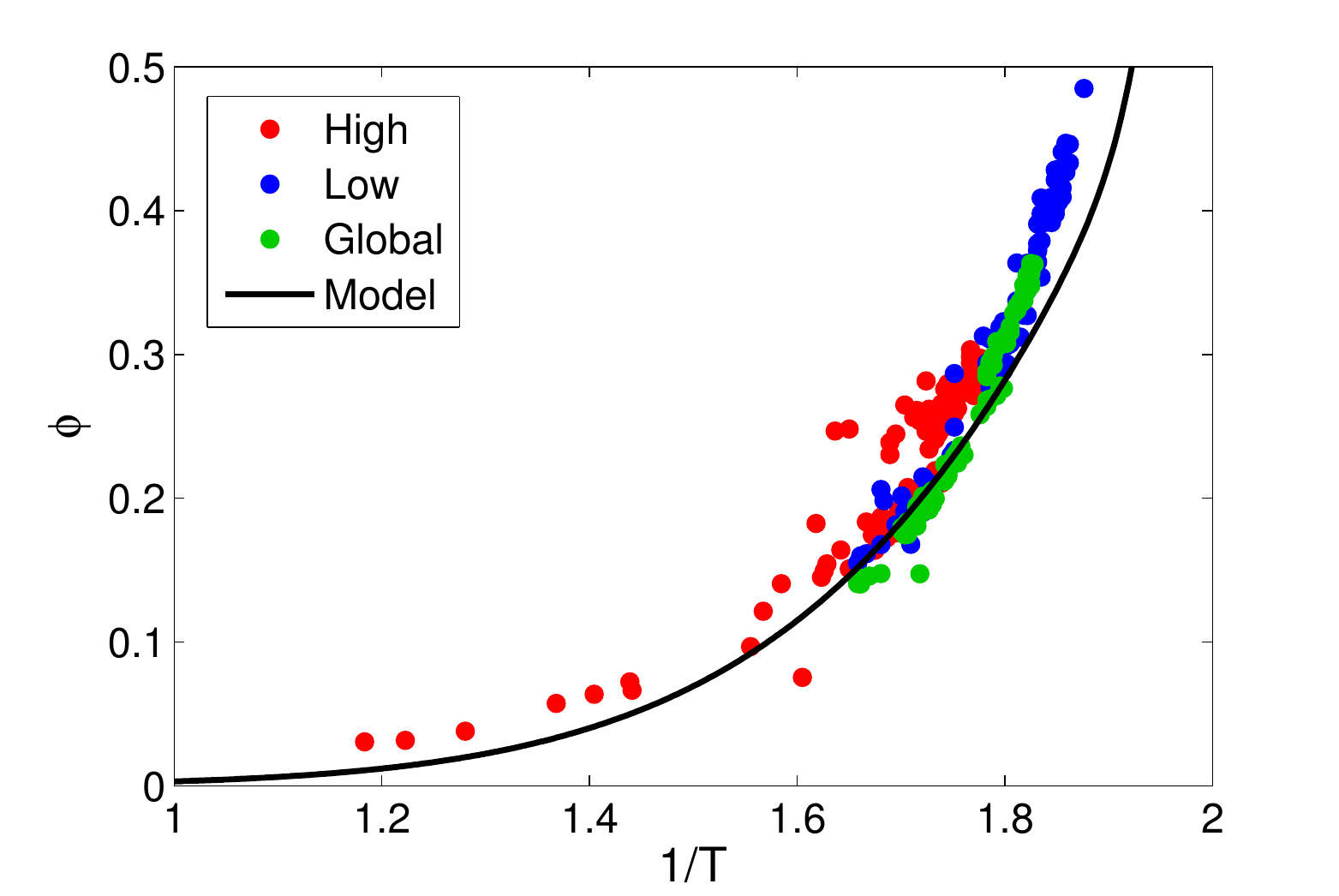}
\caption{The fitted effective temperatures and their corresponding observed values of $\phi$. The model $\phi(T)$ was formulated using quiescent data which was only accessible in the region $1/T < 1.74$. The data points of the fitted effective temperatures follow the model $\phi(T)$ reasonably closely for all observed effective temperatures.}
\label{figSegmentPhiAndTemperatures}
\end{center}
\end{figure}

Figure \ref{figSegmentPhiAndTemperatures} shows the fitted effective temperatures compared to the observed $\phi$ values and the existing model for $\phi(T)$ from \cite{pinney2015recasting}. The data points follow the model $\phi(T)$ with reasonable accuracy. The fitted effective temperatures are higher than what the observed $\phi$ would have predicted, however, it is likely that the system size dependent mesocluster size models give effective temperatures that are too high. This is evidenced in a small number of windows where the low shear segment has been fitted with an effective temperature that is actually \emph{higher} than the global average; suggesting a possibility that the models may be biased towards higher temperatures, although quantifying this bias would prove challenging.

\begin{figure}
\begin{center}
{\includegraphics[width=\columnwidth]{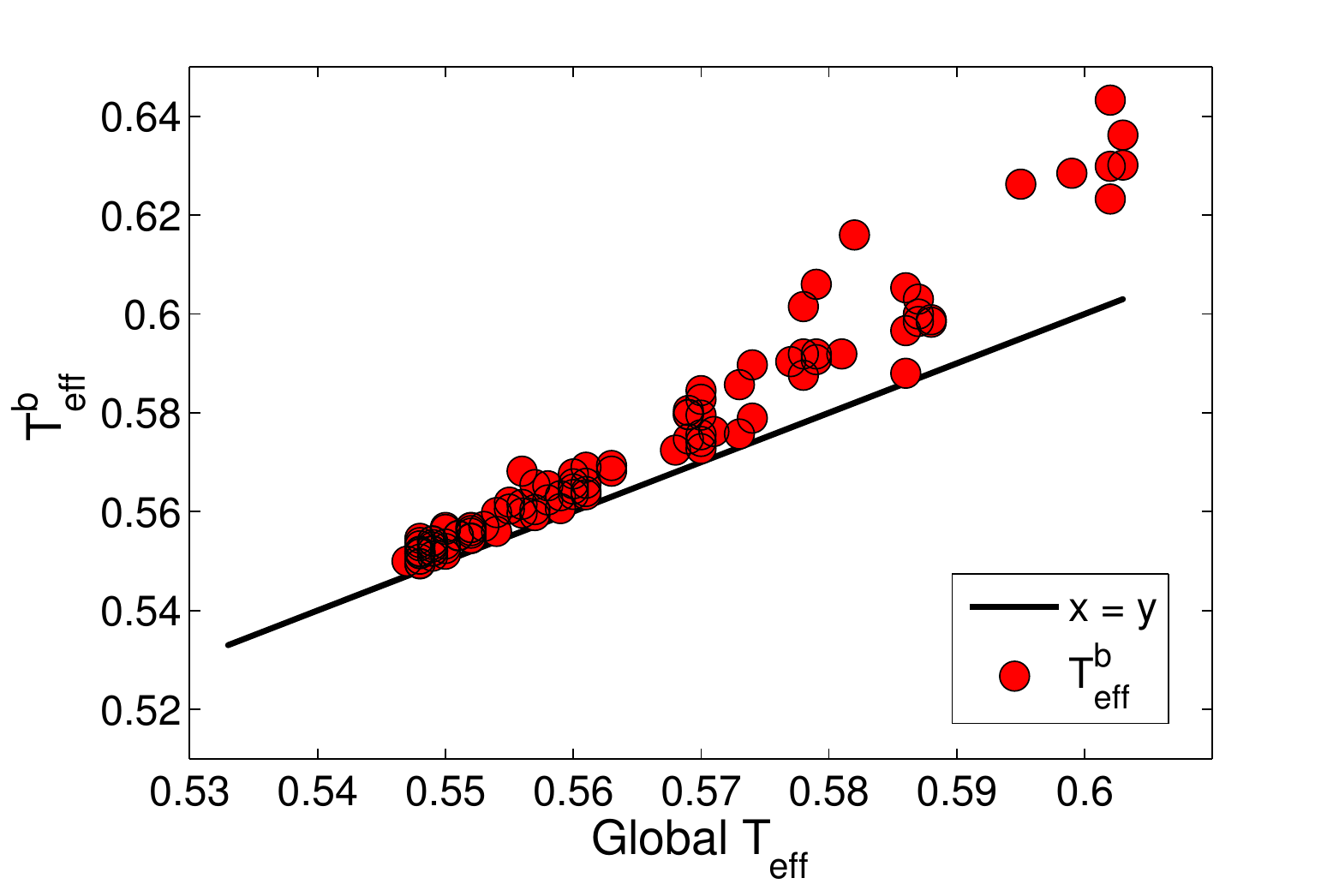}}
\caption{The global effective temperature estimated from a simple linear combination of high and low shear segments (Eq. \ref{eqSegmentModel}).}
\label{figSegmentModel}
\end{center}
\end{figure}

Using these fitted effective temperatures, we can formulate a very simple linear combination of the high and low shear segments to estimate an average effective temperature, $T^{\mathrm{b}}_{\mathrm{eff}}$ for the banded system. Note that we have neglected the ``interfaces'' between the banding and non-banding regions since they form a relatively small part of the system. Specifically:
\begin{equation}
T^{\mathrm{b}}_{\mathrm{eff}} = \frac{T_{\mathrm{eff}}(h)L(h) + T_{\mathrm{eff}}(l)L(l)}{L(h) + L(l)}
\label{eqSegmentModel}
\end{equation}
\noindent
where $L(h)$ and $L(l)$ are the sizes of the high and low segments as a proportion of the box height (e.g. 0.5 for half box height). The results for this are shown in Fig. \ref{figSegmentModel}. Given the simplicity of this linear combination and the potential combined inaccuracies from the system size dependent mesocluster size models, the estimated global effective temperatures are in reasonable agreement with the expression.

We therefore have two models describing the effective temperature(s) of the system. Taking a global view of the system, the effective temperature is determined by a linear relationship with the shear rate (Eq. \ref{eqShearTempConvergence}). If shear banding is exhibited, a local analysis shows that the system forms two distinct regions of high and low shear (determined by the local $D^{2}$ values). Hence taking account of shear banding, the effective temperatures of these regions and their relative sizes can be used to estimate the global effective temperature of the system (Eq. \ref{eqSegmentModel}). The combination of Eqs. \ref{eqShearTempConvergence} and \ref{eqSegmentModel} gives a complete description of the mesocluster size distributions expected for any simulation temperature and/or shear rate, including the segment differences which would be observed in systems exhibiting shear banding.

Before closing, we discuss the relevance of our findings in the context of shear banding. When a system undergoes shear banding, one expects that the bands have differing rigidities. Given that quiescent vitrification involves a change in rigidity, it is natural to expect that some properties of the quiescent case may carry over to the shear banding case. This is what we indeed find. In particular, a drop in the population of icosahedra in the shear bands seems entirely consistent with the idea that icosahedra are involved in the increased rigidity of the Wahnstr\"{o}m model. This is also consistent with the negative correlation identified between locally weak ``soft spots''  and icosahedra in a metallic glassformer \cite{ding2014} and banding behaviour  \cite{ding2012,feng2015}. Fitting the shear bands with our population dynamics model \cite{pinney2015recasting} suggests that the banding regions can be treated as if they are at a higher effective temperature. Interestingly, other work also correlates sheared systems with higher temperature, both in simulation \cite{nicolas2016,bailey2006} and also in experiment \cite{lewandowski2006}. Our analysis thus forms a structural connection for these observations of the relationship between temperature and shear. 

,

\section{Summary and Discussion}
\label{sectionSummary}

In this paper, we analyzed the Wahnstr\"{o}m binary Lennard-Jones model under different rates of shear for a wide range of temperatures. The system under was sheared for long enough to reach the steady state (i.e. steady stress had been achieved) before obtaining data. In this way, we were able to access temperatures inaccessible to quiescent systems. Additionally, increasing the rate of shear can be shown to act like increasing the temperature of the system. This was evidenced in the mesocluster size distributions. In particular, once the shear rate is slow enough ($\dot{\gamma} \lesssim 0.01/\tau_{\alpha}$), the system shows no obvious behavioural differences from their corresponding quiescent systems.

We conclude that (\emph{i}) at the level of our mesocluster model, shearing may be regarded as equivalent to changing temperature; (\emph{ii}) shear behaviour in regions poor in icosahedra provides strong evidence that icosahedra-rich regions are more rigid. This suggests that the formation of icosahedra may be related to local rigidity in the Wahnstr\"{o}m model. Such behaviour has been noted in metallic glasses \cite{ding2014}. Shear banding leads to two different effective temperatures, which approximately obey a simple linear superposition to the global effect temperature.

The sheared systems were initially fitted with an effective temperature using the mesocluster size distribution model from \cite{pinney2015recasting}. From this, we were able to identify an effective temperatures as a function of shear rate. For $0.56 \leq T \leq 0.8$, the effective temperatures are well described by a linear function of the shear rate (Eq. \ref{eqShearTempConvergence}). This model is only relevant in the regime where $\tau_{\alpha}$ can be evaluated. However, even in the low temperature simulations ($T \leq 0.5$), decreasing the shear rate resulted in decreasing the effective temperature of the systems. In these low temperature simulations, the observed effective temperatures were significantly colder than can be obtained in equilibrated quiescent systems. Thus, this method allows us to probe deeper into the energy landscape than can be achieved with quiescent simulation.

Many of the state points studied here exhibited shear banding. These have two distinct regions; one with a low shear rate, and the other with high shear rate. The high and low shear rate regions were identified using the non-affine deformation parameter, $D^{2}$, which measures the relative movement of neighbouring particles compared to a central one \cite{falk1998dynamics}. Higher $D^2$ values identify regions of high mobility (high shear rate) while lower values identify regions of low mobility (low shear rate). To analyze the banding, the average $D^{2}$ values and average icosahedra density in the $y$ direction was considered. A very strong negative correlation coefficient was found between the values of $D^{2}$ and density of icosahedra, suggesting a measurable difference in the mesocluster size distributions between the high shear and low shear regions. We used the $D^{2}$ values to construct boundaries for the high and low shear regions, thus allowing us to partition the simulation box into segments according to their local shearing behavior. Mesoclusters inside each of these segments were identified, and the size distributions were calculated for each high and low shear segment across all banded systems. In all cases, the low shear regions had significantly lower effective temperatures than the high shear regions.

Using a linear combination of the effective temperatures of the high and low shear segments, we could estimate the global effective temperature with reasonable accuracy (Eq. \ref{eqSegmentModel}). This result means that, given we know the global effective temperature (which can be predicted from the linear relationship between shear rate and effective temperature Eq. \ref{eqShearTempConvergence}) and the approximate size of the shear bands, we can estimate the effective temperatures of the shear bands, and vice versa.

Our work opens a perspective of using a shear to probe deep in the energy landscape, beyond the regime accessible to conventional simulation. This is made under the assumption that the properties of the mesocluster model (icosahedra population and mesocluster properties) accurately represent the system at low temperature \cite{pinney2015recasting}. In the future, this method can be generalised to system with other LFS such as the Kob-Andersen model \cite{coslovich2007understanding,malins2013fara} and hard spheres \cite{royall2015} and indeed to practical materials with well-defined LFS such as metallic glasses \cite{cheng2011}.

\subsection*{Acknowledgements} 
The authors would like to thank Andrea Cavagna, Daniele Coslovich, Jens Eggers, Bob Evans, Rob Jack, Gilles Tarjus and Francesco Turci for many helpful discussions.
CPR would like to acknowledge the Royal Society for financial support and the European Research Council under the FP7 / ERC Grant agreement n$^\circ$ 617266 ``NANOPRS'', and Kyoto University SPIRITS fund. RP is funded by EPSRC grant code EP/E501214/1. This work was carried out using the computational
facilities of the Advanced Computing Research Centre, University of Bristol.


\end{document}